\documentclass{edm_template}
\usepackage{float}

\usepackage{tabularx, colortbl}
\usepackage{listings}
\usepackage{booktabs}
\usepackage{rotating}
\usepackage{multirow}
\usepackage{adjustbox}
\usepackage{graphicx}
\usepackage{flushend}
\usepackage[stable]{footmisc}
\usepackage{hyperref}
\usepackage{makecell}

\usepackage[moderate]{savetrees}
\begin{document}

\definecolor{tablebar}{rgb}{0.6, 0.6, 0.9}
\setlength\arrayrulewidth{1pt}\arrayrulecolor{tablebar}
\title{Predicting Student Performance Based on Online Study Habits: A Study of Blended Courses}

\numberofauthors{1} 
\author{
\alignauthor
Adithya Sheshadri, Niki Gitinabard, Collin F. Lynch, Tiffany Barnes, \& Sarah Heckman\\
       \affaddr{North Carolina State University, Raleigh, NC, USA}\\
       \email{\{ aseshad, ngitina, cflynch, tmbarnes, sarah\_heckman \}@ncsu.edu}
}
\maketitle
\begin{abstract}

Online tools provide unique access to research students' study habits and problem-solving behavior. In MOOCs, this online data can be used to inform instructors and to provide automatic guidance to students. However, these techniques may not apply in blended courses with face to face and online components.  We report on a study of integrated user-system interaction logs from 3 computer science courses using four online systems: LMS, forum, version control, and homework system. Our results show that students rarely work across platforms in a single session, and that final class performance can be predicted from students' system use.     
\end{abstract}


\maketitle

\section{Introduction}

Today, students enrolled in university courses, particularly those in STEM disciplines, supplement or even supplant class attendance with online materials such as Learning Management Systems (LMSs), discussion forums, intelligent tutoring systems, automatic graders, and homework helpers.  TAs now offer online office hours even for local students and lecture videos are reviewed in lieu of note-taking. The goal of these tools is to foster a rich learning environment; to support good study habits; and to enable instructors to give and grade realistic activities such as collaborative learning and project development \cite{graham06}. In addition to planning class lectures, course instructors now manage a constellation of online services which students can use, or not, at their own pace. In practical terms, many traditional face-to-face on campus courses are blended courses.  

While these tools can be beneficial, blended courses normally use several different tools, requiring students to switch between different websites several times to access lecture notes, online discussions, and assignment submission systems. Even when the tools are linked through a single LMS such as Moodle, the need to transition from platform to platform can be challenging.  In order to engage effectively with such hybrid materials, students need to develop good online study skills and need to effectively integrate information across platforms.  While we have generally assumed that technically-savvy students have these skills, little work has been done to empirically evaluate how students use these platforms and whether or not their observed habits can be used to predict their performance.

While many of the systems used in blended courses are similar to those in Massive Open Online Courses (MOOC), studies of blended courses are limited.  Prior work has mostly focused on single tools, and generally lack early prediction power \cite{zafra12, romero2013web, park2017, agnihotri2015mining, gitinabard17w, ruizpredicting}. Prior research has suggested novel methods that can be used to predict student performance in MOOCs, based upon features extracted from students' interactions with different learning materials \cite{Brooks2015, sinha14, li17, kloft14}. Since not  all of the students' learning activities can be monitored online, it is not certain whether the same methods can be applied \cite{an2017}.

In this paper, we present our work on the automatic analysis of students' study behaviors in blended courses.  We focus on 3 Computer Science courses at North Carolina State University using 4 online platforms: Moodle, a large-scale LMS; Piazza, an online discussion forum for student questions; Github Enterprise, a project management platform for software development; and WebAssign, an online homework and automatic grading platform. Our goals in this work are: to \emph{synthesize} data from these heterogeneous learning platforms; to \emph{extract} meaningful student behaviors from the interaction data; and to \emph{model} students' behaviors to predict their future performance and thus to provide a basis for automatic feedback or instructor support.  We want to leverage this synthesized data to analyze not only \emph{what} features of these online platforms students use but \emph{when} they do so.  This, in turn, can give us a deeper understanding of students' study habits and allow us to distinguish effective strategies from ineffective ones, facilitating automated support.  

We address the following research questions: (Q1) Do students focus on one tool at a time or work across platforms?; (Q2) Do students' online study habits follow clear patterns based upon the course deadlines?; and (Q3) Can students' study habits be used to predict their final scores?  We hypothesize that students tend to \emph{silo} their work in one platform at a time, and that their tool use will predictably follow course deadlines.  We also hypothesize that these patterns, or deviations from them, can be used to predict their overall performance across classes.

\section{Related Work}
The use of online tools such as discussion forums, learning management systems (LMS), and online assignment submission systems in classrooms can provide researchers with more information on students' behavior than they can obtain through observation alone.  MOOCs are attractive to researchers in part because they provide a \emph{data choke-point} which highlights most relevant student interactions.  While most of the data available in MOOCs is also available in blended courses, many of the findings on MOOCs have not been replicated in blended (face to face and online) courses in part due to a lack of available datasets and the fact that not all learning activities are tracked or logged.  In this section, we first discuss some of the studies on MOOCs and performance prediction, then we discuss the research on blended courses.

There have been a number of studies of students' behavior in MOOCs and whether or not it is correlated with their overall performance. Seaton et al., analyzed students' use of course materials on an existing EdX MOOC with the goal of determining when and how the students attempt assignments and how they divide their time over the differing activities \cite{seaton2014does}, finding that only 25\% of the students attempt more than 5\% of the assignments but those students account for 92\% of the total time spent on the course. Sixty percent of the total time spent on the class was invested by the 6\% of students who earned a certificate. There have been several attempts to predict student certification and dropout in MOOCs using features extracted from their online activities such as the number of videos watched in a week, the number of quiz or assignment attempts, the number of forum posts made per week, post length, relative time spent on assignments, and so on \cite{pursel2016, fei15,andres16, chen17}. Some researchers have gone further by defining study sessions for students, and using the sequence of students' access to the online material to make predictions on performance \cite{amnueypornsakul14, Brooks2015, li17}. Amnueypornsakul et. al. defined active study sessions for each student and used the sequence of actions within sessions to define features such as length of the action sequence, the number of occurrences of each activity, and the number of Wiki page views \cite{amnueypornsakul14}. Li et. al. applied the idea of N-grams to the sequence of student actions in a session and used those N-grams to predict the students' certification \cite{li17}. Brooks et. al. defined sessions with fixed duration such as 1 day, 3 days, 1 week, and 1 month throughout the semester and recorded students' activity within each time unit as a binary feature \cite{Brooks2015}. They defined N-grams on the sequences of features to make early and cross-class predictions of student dropout. Sinha et. al. added the concept of an interaction graph which connected students to the resources they accessed and found that a predictive model trained on the graph features can outperform N-gram based models for student behaviors \cite{sinha14}. 

These prediction methods, if applicable to blended courses, could help instructors to identify struggling students early in the semester for support.  But it is still not clear how or if these behaviors can transfer from one domain to another. An et. al. tried replicating some of the predictive methods found in MOOCs to a blended course and found that those findings can be applied with some caveats as there were some changes needed in the design process for it to be applicable in other contexts \cite{an2017}. For example, students' download activities were shown to cluster students into two categories of completing and auditing both in a MOOC and an online course, but this pattern was not visible in the blended course of their study. However, this clustering based upon assessment scores could identify some of the groups visible in MOOCs, also in the blended course.

Prior analyses of student behaviors in blended courses show that the overall level of activity increases when exam, quiz, or assignment deadlines are near at hand \cite{park2017}. Analyses of students' login behaviors also show that the students' activities follow a predictable weekly cycle dropping on Saturdays and then rising on Sundays as they prepare for the week ahead \cite{agnihotri2015mining}. Research has also shown that better performing students usually start and end their activities earlier than their lower performing peers \cite{willman15} and that it is possible to use some student activity features to predict their performance \cite{zafra12, romero2013web, gitinabard17w, agnihotri2015mining, Spacco15}. Some examples of these features are attendance, emotions during lecture, number of assignments done, the time they took to do those assignments, number of posts on the discussion forum, prior scores, number of attempts, etc.  Most of this prior work has been based upon \emph{complete} datasets which represent all of the information obtained during a semester.  Such models cannot therefore be used for early warnings or interventions. Munson et. al., by contrast, showed that features such as early scores, hours coding, error ratio, and file size can identify struggling students in the first three weeks of the class \cite{munson18}.

We focus on evaluating students' online tool use in terms of \emph{sessions}, consecutive sequences of study actions that occur between breaks for food or sleep.  Sessions have been previously used to analyze student behavior in MOOCs and in other cohesive online tools such as an LMS (e.g. \cite{Kovanovic:Penetrating:2015}).  Our work here extends that research by applying to heterogeneous data from blended courses where students can work across platforms and over longer periods of time with the goal of developing early predictors that can be used to identify high- or low-performing students in time for an intervention.  

\section{Datasets}

In order to address our research questions, we collected student data from three blended courses in Computer Science offered at North Carolina State University. Two are offerings of ``Discrete Mathematics for Computer Science,'' (DM) from 2013 and 2015 respectively.  The other was an offering of ``Programming Concepts in Java'' (Java) from 2015.  Both courses are core courses for students majoring in CS and both are structured as blended courses.  Students in both DM and Java use Moodle, an open-source LMS that is used for all courses at NC State University, and Piazza, an on-line Q\&A platform for question answering that can be used for threaded discussions. The students in the DM classes use WebAssign and the students in the Java class used Github for assignment submissions.

We collected data from Piazza and Moodle for all these classes as well as data from the WebAssign system for the 2013 DM course.  We also collected a complete record of students' Github commits from the 2015 Java course.  The data was collected as web logs, database dumps, and in the case of WebAssign, via screen scraping.  We then cleaned up the raw data, linked it across platforms, and anonymized it to produce a single coherent database for analysis.

For the purposes of this analysis, we focused solely on the students' actions and ignored actions by the course instructors.  We also eliminated students who dropped out of the course (only possible during the first two weeks of the course) as well as students who did not get a grade (at most 20 students per class).  Table \ref{tab:sources} shows the total number of actions recorded for each tool along with the average number of actions per student removing zero values.  

\begin{table}
\begin{center}
\begin{adjustbox}{width=0.40\textwidth}

\begin{tabular}{l|lccc}
\hline
Class & Source & Total actions & Avg. per student & $\sigma$ \\
\hline
DM 2013 & Moodle & 17148 & 166 & 88\\
 & Piazza & 2557 & 15 & 28\\
 & WebAssign & 265510 & 1062 & 201\\
\hline
DM 2015 & Moodle & 21972 & 80 & 59\\
& Piazza & 2208 & 12 & 17 \\
\hline
Java 2015 & Moodle & 101180 & 613 & 266\\
& Piazza & 2556 & 20 & 25\\
& Github & 31438 & 196 & 140\\
\hline
\end{tabular}

\end{adjustbox}
\end{center}
\caption{Actions Taken by Platform}
\label{tab:sources}
\end{table}

\begin{figure}[ht]
\centering
\includegraphics[width=0.35\textwidth]{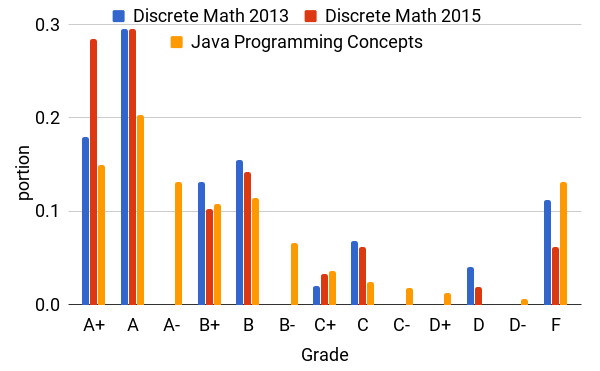}
\caption{Grade Distribution for the Three Classes}
\label{fig:grades}
\end{figure}

The final grades of all the students were provided to us by the instructors. In each case, the Grades are represented ordinally from \textbf{A$+$} to \textbf{F} from left to right. The relative distribution of student grades in the courses are shown in Figure \ref{fig:grades}.  A majority of the students achieved good grades in each course offering (i.e. B or above) and fewer than 16\% of them failed. We therefore based our predictive models on distinguishing students who achieved distinction in the course (A- to A+ grades) from those who did not.  Due to the high proportion of distinction this yielded nearly balanced datasets.

\subsection{Discrete Mathematics}
Discrete Mathematics (DM) is a core CS course that introduces students to the basics of logic and proofs, set theory, and probability.  Students typically enroll in the course during their second year and must pass it with a C or better to progress in the curriculum. Our data comes from the Fall 2013 and Fall 2015 offerings of the course.  In both cases, the course was broken into two sections with two instructors and five shared Teaching Assistants (TAs).  The 2013 offering had 250 students enrolled  while the 2015 offering had 277.  The average final grades in 2013 and 2015 were 89.5 and 86.9 out of 100 respectively.  In both years the course had 10 homework assignments (30\%), 5 lab assignments (10\%) and 4 tests accounting for 60\% of their final grades. 

Students in the DM courses used WebAssign, an online homework platform. It is used to deliver, view, and grade student assignments. Assignments may be graded both manually and automatically. WebAssign is used in the DM course for weekly assignments linked via Moodle. We have access to WebAssign data for 2013 DM offering. Each submission shows a single attempt to answer a question in an assignment and provides information on the student making the submission, the time of the submission, the assignment, question information and the sub-question part being completed.

\subsection{Programming Concepts in Java}
Programming Concepts in Java is also a core CS course that covers software topics such as: system design and testing, encapsulation, polymorphism, finite-state automata, and linear data structures and recursion. Like the DM course it is offered to students during their second year and students must pass with a C or better.  To obtain the letter grade earned, the students must obtain an average of 60\% or better on the exams and assignments. In 2015, this class was structured into three sections with one instructor per section.  One  section was a pure distance education section with a completely separate student population.  This was omitted from our analysis for the sake of consistency. Our dataset, therefore, covers 2 sections with 2 instructors, 9 teaching assistants, and 180 local students.  The high TA to student ratio is due to the fact that this course involve a substantial coding project components and also the external funding supporting additional TAs.  The course included 3 projects, 2 midterms, and one final exam and their final grades are generated based on the grades on all these activities. The average final grade for this class was 79.7.  

Students in the Java course use GitHub, a version control system used widely in Software Engineering projects to allow users track changes to the code and share coding tasks within a team. Github is used in the Java class as a tool for individual and team projects, and also as an assignment submission system. The system is connected to an automatic test suite and students are graded based on their latest Github commits. Each record in our logs identifies the author, the number of changes to the code, and the time of submission.

\section{Sessions}
We aggregated the heterogeneous actions described above into a unified transaction log. 
This data consists of 285,465 transactions from the DM 2013 class, 24,180 transactions from the DM 2015 class, and 135,351 transactions from the Java class. As Table \ref{tab:sources} shows, the lion's share of these transactions are WebAssign actions from Discrete Math 2013 and Moodle actions from the other two course offerings.

We divided the individual transactions into \emph{sessions} representing contiguous sequences of student actions using data-driven cutoffs. 
Our goal in grouping the student actions was to better understand the students' online study habits and their longer-term strategies. Aggregating student actions into sessions is a nontrivial problem.  Fixed time cutoffs can have incorrect edge cases and the time cutoff used can affect our final results.  Kovanovic et. al., for example, evaluated the impact of different time cutoff strategies for a dataset extracted from a single LMS \cite{Kovanovic:Penetrating:2015}. They found that there was no one best cutoff and recommended exploring the data to select a context-appropriate cutoff. Some techniques that have been used to estimate sessions are:\\
\-\hspace{0.5cm}\textbf{Fixed duration:} Sessions can be defined based upon a fixed (often a priori) unit of time such as an hour or a day as in \cite{Brooks2015}.  Sessions can also be defined by considering periods between assignment deadlines as the duration of the session.\\
\-\hspace{0.5cm}\textbf{Cutoff:} Another method is using a fixed timeout or cutoff similar to Amnueypornsakul et. al. \cite{amnueypornsakul14}.  Here we group student actions into sections and separate sections when they go offline or pause for a set period of time. The consecutive actions between pauses then belong to a single session irrespective of its duration. \\
\-\hspace{0.5cm}\textbf{Navigation:} In this approach, common in web-analysis, the actions themselves are analyzed to determine when a session has ended.  If, for example, a user traverses to an unrelated content or engages in off-task browsing then we consider the session to be over.

\begin{figure}[t]
\centering
\includegraphics[width=0.35\textwidth]{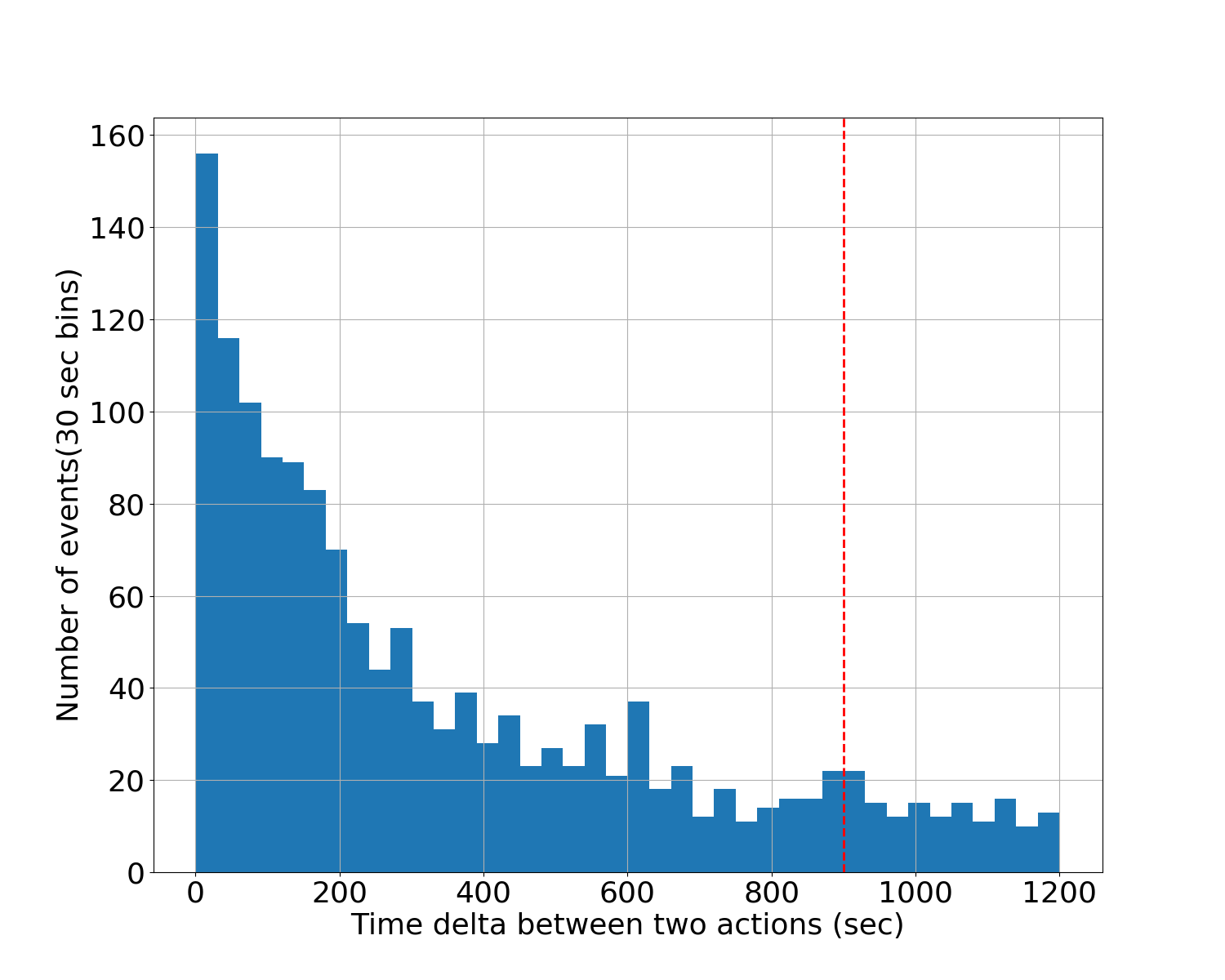}
\caption{Histogram for Gap Length with Change of Platform for the 2013 Discrete Math Class.}
\label{fig:sessionsChange}
\end{figure}

Given the variation of the data, it is not possible to define a single cutoff value that accurately captures all cases. We, therefore, opted to define two session types with two different cutoff values.\\
\-\hspace{0.5cm}\textbf{Browser Session: } $m=15$ minutes indicating a short break likely with the same browser open.\\
 \-\hspace{0.5cm}\textbf{Study Session: } $m=40$ minutes indicating that student likely changed tasks or quit working entirely.

The \emph{Browser Session} can be characterized as a case where the students are actively working on a single task with little change. This may include working on a multi-part WebAssign question, reading through materials on Moodle, or diagnosing an issue with their code with guidance from Piazza. Sessions of this type were comparatively short in duration.  The \emph{Study Session} by contrast allows for a much larger gap where students may shift from reading materials to answering questions or engaging in (online) discussions with their peers, and back again. This large cutoff was based upon the cross-platform breakdown and was in part intended to address our lack of data regarding the students' offline activities.

\section{Results \& Analysis}

Table \ref{tab:session_hvh} presents basic statistics on the two types of sessions across the three classes. Because these sessions are defined by a time cutoff they have overlapping instances. Thus in DM 2013, 12,349 of the sessions were both study and browser sessions meaning that the gap between them and the neighboring sessions was over 40 minutes long and all of the internal gaps were less than 15 minutes. When analyzing the session duration, we found that almost half of the sessions (of both types) were comparatively short with the students making less than five actions. The relative frequency of the sessions drops quickly as the session length increases.  Similar trends were exhibited when we examine the length of each session based on their duration. We also observe that the average duration of the sessions in DM 2015 dataset was drastically shorter than those in DM 2013 and Java 2015. This may be explained by the fact that the WebAssign records were present in DM 2013 while GitHub was included in Java 2015.  This would give a more frequent and detailed picture of students' problem-solving. Removing Github activities from the Java class records resulted in a similar pattern of shorter sessions.

\subsection{Q1: Homogeneity}

The browser and study sessions can be classified as heterogeneous and homogeneous sessions.  \emph{Heterogeneous sessions} occur when the student switches between platforms at least once during the session.  \emph{Homogeneous sessions} occur when no such change takes place. Table \ref{tab:session_hvh} presents a breakdown of the two types across the classes. As the table illustrates, in all of the classes more than 95.5\% of the browser sessions are homogeneous as are more than 93.8\% of the study sessions.
These results are consistent with our hypothesis that students do not work across platforms but instead \emph{silo} their activities working on one system at a time. This is true even with the long timeout for the study sessions. When they do transition from one platform to another it is largely a transition between Moodle, which links course schedules to assignments, and WebAssign, which allows them to complete their assignments in the DM 2013 class, or between  Moodle and Github in the Java class.

\begin{table}[t]
\begin{center}
\begin{adjustbox}{width=0.47\textwidth}
\begin{tabular}{l|lcccc}
\hline
Class & Session & Count & Avg   & Homogeneous & Heterogeneous\\
& & & Duration & & \\
\hline
DM 2013 & Browser  & 17699 & 9 min & 16892 & 777\\
& Study & 14574 & 16 min & 13668 & 906 \\
\hline
DM 2015 & Browser & 10981 & 2 min  & 10963 & 18\\
& Study & 10038 & 2 min & 9994 & 44\\
\hline
Java 2015 & Browser & 28768 & 5 min  & 28645 & 123\\
& Study & 25005 & 17 min & 21932 & 223\\
\hline
\end{tabular}

\end{adjustbox}

\caption{Information on Different Types of Sessions}
\label{tab:session_hvh}
\end{center}
\end{table}

\subsection{Q2: Patterns}

\begin{figure*}
\centering

  \includegraphics[width=\textwidth]{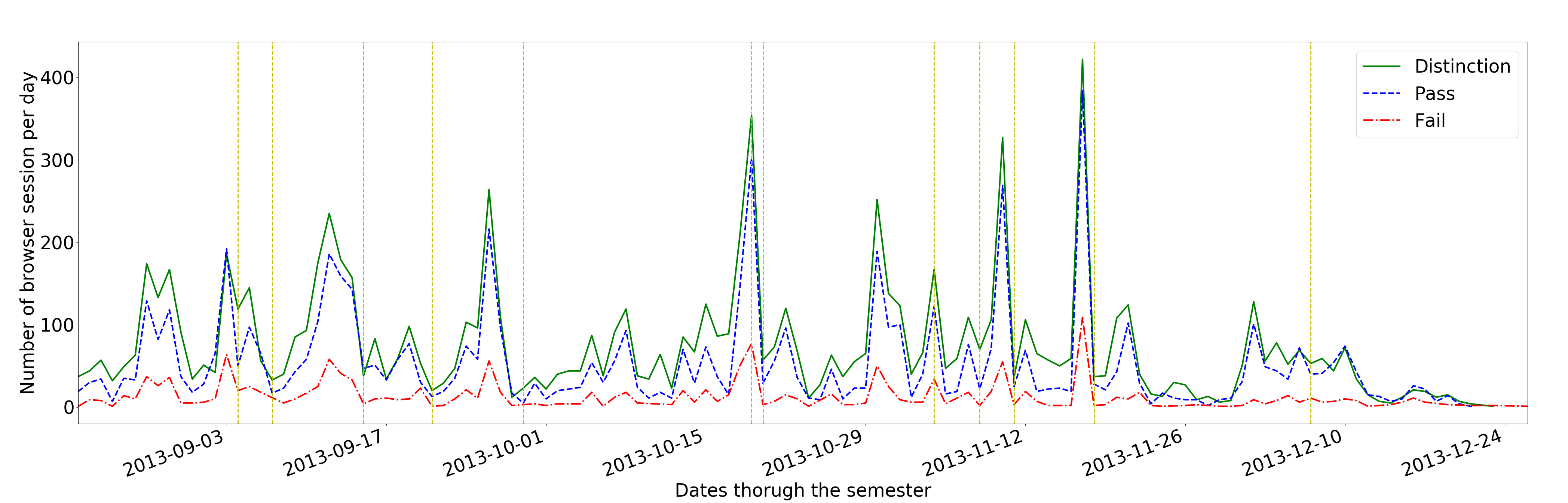}

\caption{Frequency of Browser Sessions with Assignment Deadlines and Test Dates for Discrete Math 2013}
\label{fig:sessionOccurancesTA_DM13}
\end{figure*}

As noted above, the grade distribution for the courses is quite high. We therefore classified the students into one of three categories for analysis: \emph{Distinction} (A+, A, or A-); \emph{Pass} (B+, B-, B, C+, or C-); and \emph{Fail} (D or F).  We plotted the the frequency of individual browser sessions day by day over the course of the semester, and example of them for DM 2013 is shown in Figure \ref{fig:sessionOccurancesTA_DM13}. The red line indicates the Fail group, the blue line corresponds to Pass group and green represents the Distinction group. The vertical bars show the due dates for assignments and exams.  As the plot illustrates, the number of sessions spike before each deadline for \emph{all three} of the performance groups. A similar pattern was observed for the other classes, the frequency of the study sessions and also for the duration of both session types sessions. These results are consistent with our hypothesis and prior studies that students are deadline-driven even in blended courses. We also observe that the Fail group performs much fewer activities than the other two groups, getting close to zero in DM 2015 class. This shows that most of the actions the Fail group performed were WebAssign actions, which we do not analyze for the 2015 class. 

\begin{table}
\begin{center}
\caption{Kruskal-Wallis P-values for Succeed/Fail Classification, Values Less than 0.05 Are Illustrates in Bold}
\label{Tbl:Kruskal-Results}
\begin{adjustbox}{width=0.47\textwidth}

\begin{tabular}{l|ccc|ccc|ccc}
  
\hline
  & \multicolumn{3}{l|}{\textbf{Pre Test 1}} & \multicolumn{3}{l|}{\textbf{Pre Test 2}}  & \multicolumn{3}{l}{\textbf{Full Semester}}\\

Parameter & DM  & DM  & Java  & DM  & DM  & Java  & DM  & DM  & Java \\
& 2013 & 2015 & 2015 &  2013 & 2015 & 2015 & 2013 & 2015 & 2015 \\
\hline
Avg Gap & 0.15 & \textbf{0.02} & \textbf{0.00} & \textbf{0.00} & \textbf{0.03} & 0.12 & 0.10 & \textbf{0.04} & \textbf{0.02}\\
Num Sessions & \textbf{0.01} & \textbf{0.03} & \textbf{0.00} & \textbf{0.00} & \textbf{0.02} & 0.06 & \textbf{0.00} & \textbf{0.01} & \textbf{0.04}\\
Pratio & \textbf{0.00} & \textbf{0.00} & \textbf{0.00} & \textbf{0.00} & \textbf{0.00} & \textbf{0.04} & \textbf{0.00} & \textbf{0.00} & 0.08\\
Total Time & \textbf{0.00} & 0.13 & \textbf{0.04} & \textbf{0.00} & 0.23 & 0.25 & \textbf{0.01} & 0.20 & 0.08\\
Consistency & 0.07 & \textbf{0.01} & \textbf{0.00} & \textbf{0.00} & \textbf{0.03} & 0.10 & 0.08 & \textbf{0.03} & \textbf{0.03}\\
Total Actions & \textbf{0.00} & 0.14 & \textbf{0.02} & \textbf{0.00} & 0.08 & 0.14 & \textbf{0.01} & 0.17 & 0.19\\
Piazza Questions & \textbf{0.00} & \textbf{0.00} & \textbf{0.04} & \textbf{0.00} & \textbf{0.00} & 0.08 & \textbf{0.00} & \textbf{0.00} & 0.10\\
Piazza Answers & \textbf{0.00} & \textbf{0.00} & \textbf{0.04} & \textbf{0.00} & \textbf{0.00} & 0.09 & \textbf{0.00} & \textbf{0.00} & 0.11\\
MultipleAttempts & \textbf{0.00} &  &  & \textbf{0.00} &  &  & \textbf{0.00} &  & \\

\hline
\end{tabular}

\end{adjustbox}

\end{center}
\end{table}
\subsection{Q3: Prediction}

As Figure \ref{fig:sessionOccurancesTA_DM13} and the other class plots illustrated, all three performance groups in all three classes followed a similar pattern. All of the groups have irregular activity patterns and all of them see spikes prior to each of the deadlines and exams. Yet there are important differences among the groups.  As a group, the Distinction students were always active, with the number of active sessions rarely if ever, reaching zero. The Fail students, by contrast, were frequently inactive as a group with long periods where no fail student was active at all. The Pass students, by contrast, occupy an interesting middle ground with less consistent activity than the Distinction group but far more than the Fail group. This suggests that students who succeed in the courses are broadly more consistent and engage in online activity at regular intervals.  However, all the students are given to cramming (spending much more time on classes right before tests) with the better students cramming as much or more than their peers. These results also indicate that the relative gap between sessions may be a significant predictor of students' individual performance.

Based upon those results we then identified a set of 16 session features for deeper analysis: Sessions: count, avg actions, total actions; avg duration, time=count* avg duration, avg gap,  consistency, \#homogeneous, \# heterogeneous; Piazza: \#questions, \#answers, ratio of sessions containing Piazza activity; Webassign (DM2013 only): \# parts submitted, \# first attempts, \# multiple attempts; Performance (distinction, pass, fail);

\vspace{-1em}
\begin{displaymath}
 Consistency= AvgGap * (max(Sessions Count) - Sessions Count)
 \end{displaymath}


We reclassified the students into two categories, \emph{Distinction} and \emph{Non-distinction} (Fail and Pass students) and applied the Kruskal-Wallis (KW) test, a non-parametric analogue to the Analysis of Variance test (ANOVA) \cite{kruskal1952use}, to determine whether or not any of these features are significantly correlated with having high performance in the course. The Kruskal-Wallis test is a good choice in this context because it does not assume normally-distributed data. Table \ref{Tbl:Kruskal-Results} lists the features that were significant among the groups. As that table illustrates 9 of the features were statistically significant predictors of whether or not the students would pass the course.  Crucially, some of these features were significant predictors of student performance even when we restricted our focus to data from the first half of the semester or to the first quarter (3 weeks). It is not surprising that the significant features differed between the classes given the absence of WebAssign data from two of the courses and the use of GitHub in the Java class.  It is interesting however, that even without including WebAssign data source in the Discrete Math 2015 class, we can observe significant correlations between the defined features and performance. Our results illustrate that in all the classes, most of these features are significant early in the semester and the sign of the coefficients do not change in different time frames and across classes.

We extended this analysis by testing whether or not these values correlated with students' final grades using Kendall's $\tau$ a non-parametric correlation coefficient that works well with small sample sizes and is robust in the presence of ties \cite{dalgaard2008introductory}. Table \ref{Tbl:Spearman} lists the $\tau$ coefficient and p-values for the features that were significantly correlated with the students' final grades. As the table illustrates, most of the features were significantly correlated with the final grades in all the classes, though the coefficients were small.  Moreover, the direction of the correlations did not change over the course of the semester. In the 2015 Java dataset however, none of the features were correlated with the data before test 2. It is not entirely clear why the results are so different for this course.  The gap may be explained by a change in the course activities in the second part of the class that is not adequately reflected in our dataset.

\begin{table*}[t]
\begin{center}

\begin{tabular}{l|ccc|ccc|ccc}
\hline
& \multicolumn{3}{l|}{\textbf{Data before Test 1}}& \multicolumn{3}{l|}{\textbf{Data before Test 2}} & \multicolumn{3}{l}{\textbf{Total Semester Data}}\\
\hline
& DM 2013& DM 2015& Java 2015& DM 2013& DM 2015& Java 2015& DM 2013& DM 2015& Java 2015\\
\hline
Avg Gap& -0.075& -0.1209**& -0.1334*& -0.1411**& -0.1171**& -0.0994& -0.0917*& -0.1212**& -0.1629**\\
Num Sessions& 0.1413**& 0.1445***& 0.1548**& 0.1833***& 0.1386***& 0.0913& 0.2121***& 0.1484***& 0.1427**\\
Pratio& 0.2461***& 0.2036***& 0.1192*& 0.2502***& 0.2196***& 0.0712& 0.3090***&  0.2299***& 0.0716\\
Total Time& 0.1638***& 0.1382***& 0.1205*& 0.1858***& 0.1205**& 0.0795& 0.1759***& 0.1210**& 0.1647**\\
Consistency& -0.0879*& -0.1253**& -0.1374**& -0.1456***& -0.1197**& -0.099& -0.0996*& -0.1233**& -0.1535**\\
Total Actions& 0.1782***& 0.1225**& 0.1154*& 0.2038***& 0.1265**& 0.076& 0.1648***& 0.1131**& 0.1319*\\
Test 1& & & & & & & 0.5141***& 0.5216***& 0.5141***\\
Test 2& & & & & & & 0.4783***& 0.6582***& 0.4783***\\
\hline

\end{tabular}
\caption{Kendall's $\tau$ for the Defined Parameters and the Final Course Outcome (*: $p<0.05$, **: $p<0.01$, ***: $p<0.001$)}
\label{Tbl:Spearman}

\end{center}
\end{table*}

\begin{table*}[t]

\begin{center}
\begin{adjustbox}{width=0.87\textwidth}
\begin{tabular}{l|ccc|ccc|ccc}
\hline
  & \multicolumn{3}{l|}{\textbf{Before Test 1}} & \multicolumn{3}{l|}{\textbf{Before Test 2}}  & \multicolumn{3}{l}{\textbf{Full Semester}}\\

Parameter & DM 2013 & DM 2015 & Java 2015 & DM 2013 & DM 2015 & Java 2015 & DM 2013 & DM 2015 & Java 2015\\
\hline
Decision Tree & 0.5432 & 0.3881 & 0.5172 & 0.6750 & 0.6032 & 0.7500 & 0.6774 & 0.6769 & 0.7273\\
KNN & 0.5067 & 0.4815 & 0.3922 & 0.5352 & 0.4444 & 0.5333 & 0.7164 & 0.7857 & 0.7778\\
Logistic Regression & 0.6364 & 0.3182 & 0.6885 & 0.6333 & 0.5763 & 0.7600 & 0.6452 & 0.8077 & 0.6939\\
\hline
\end{tabular}
\end{adjustbox}
\caption{F-Measure Performance for Distinction Group Prediction}
\label{tab:regression}
\end{center}
\end{table*}
These results are largely consistent with our hypotheses, particularly for the DM offerings. We can use individual variables to predict whether or not the students will pass the course with high performance, based upon some of their per-session features. More importantly, we can do so based on the first few weeks of the course. Thus, it is possible to identify students who may be in need of support early when there is still time to change student behaviors.  

\subsection{Predictive Models}

We expanded on these results by training predictive classifiers for the students' course performance based upon the correlated features. For the models including data after test 1, we included the test 1 grade and for the model based on all semester data, we included both tests 1 and 2 grades. We used logistic regression, decision tree, and K Nearest Neighbor classifiers to predict student performance using data generated before test 1, before test 2, and over the course of the entire semester. We generated a model for each course and subset, which could classify students into Distinction/Non-Distinction groups. The $F1$ scores for these models are shown in Table \ref{tab:regression}. While the best performing classifier varies among different classes and subsets, the best models based upon full semester or even before test 2 activities for all the classes performed with decent accuracy.

\section{Conclusions \& Future Work}

Blended courses which pair face to face lectures with multiple distinct online learning platforms are increasingly the norm, particularly in STEM domains. Our goal in this paper was to determine whether or not it is possible to automatically analyze students' online study behaviors to identify good and poor study habits with the goal of supporting instructors and of providing automated guidance. In particular, we sought to address the following three research questions: (Q1) When working with online resources do students focus on a one tool at a time or do they work across platforms?; (Q2) Do students' online study habits follow clear patterns based upon the course deadlines?; and (Q3) Can students' observed study habits be used to predict their final scores?

In order to address these questions, we collected data from three CS course offerings at North Carolina State University. Two were separate instances of the same course while the other represented a different topic and instructor.  All three courses used a range of online tools, we collected data from four critical ones: a shared LMS, an online discussion forum, an online homework platform, and a version control system. We then developed methods to synthesize this heterogeneous student data across the platforms and examined students' individual study actions grouping them into study and browser sessions using empirical cutoffs.  We then grouped students into separate categories based on their performance and analyzed the pattern of sessions observed for each group.  And finally, we identified key features of the students' online habits, assessed the relative correlation of those habits with their final performance, and trained classifiers to predict their performance.

In each case, we found that the data was consistent with our hypotheses.  Students in the course typically siloed their work on the platforms (homogeneous sessions) and rarely, if ever, used two or more platforms in a single session. The students' study and browser sessions spiked in advance of each course deadline or test and dropped precipitously afterward. This pattern was consistent for these undergraduate students regardless of their final performance..  And finally, we found that the students' study habits did differ based upon their level of performance and that key features of the study habits were significantly correlated with the students' performance and final grades.  Moreover, some of these correlations were observed even in the first few weeks of the course, at a time when change is still possible. The features identified can be used to construct successful classifiers to predict performance and the individual features (e.g. average gap between sessions), lend themselves to clear direct feedback.  

Prior researchers have shown that it is possible to analyze students' actions on MOOCs to predict their ultimate performance in the course.  In MOOCs, we have a data choke point that yields largely complete records of students' course activities.  In blended courses, however, we lack a complete data picture as the students still engage in face-to-face lectures, visit office hours, and meet directly to discuss materials, or to exchange solutions.  In spite of this incomplete information we have shown that it is possible to analyze students' behaviors to derive pedagogically relevant information that can be used to support instructors or to provide automated guidance.  While the induced classifiers are not perfect, and while they depend upon some crucial design decisions such as the session selection, they still have the potential to provide real benefits in everyday classrooms.      

In the near term, we plan to extend this work by incorporating additional data that was unavailable for our present analysis such as records from Jenkins, an automatic test suite. We also plan to investigate other mechanisms to detect off-task behavior and to estimate time on task that are sensitive to the specific actions being taken. In longer-term work, we plan to develop a centralized platform for automatically logging and integrating data from heterogeneous sources to provide automatic strategic feedback. It is our hypothesis that regular feedback from a virtual ``study buddy'' can be useful in encouraging students to develop better work habits even with a relatively low rate of guidance.

\section{Acknowledgments}
 This material is based upon work supported by the National Science Foundation under Grant No.  1418269: ``Modeling Social Interaction \& Performance in STEM Learning'' Yoav Bergner, Ryan Baker, Danielle S. McNamara, \& Tiffany Barnes Co-PIs, and by a Google CS Capacity award, and an NCSU DELTA Course Redesign Grant.

\clearpage
\bibliographystyle{abbrv}
\bibliography{Patterns}  

\begin{thebibliography}{10}

\bibitem{agnihotri2015mining}
L.~Agnihotri, A.~Aghababyan, S.~Mojarad, M.~Riedesel, and A.~Essa.
\newblock Mining login data for actionable student insight.
\newblock In {\em Proc. 8th International Conference on Educational Data
  Mining}, 2015.

\bibitem{amnueypornsakul14}
B.~Amnueypornsakul, S.~Bhat, and P.~Chinprutthiwong.
\newblock Predicting attrition along the way: The uiuc model.
\newblock In {\em Proceedings of the EMNLP 2014 Workshop on Analysis of Large
  Scale Social Interaction in MOOCs}, pages 55--59, 2014.

\bibitem{an2017}
T.-S. An, C.~Krauss, and A.~Merceron.
\newblock Can typical behaviors identified in moocs be discovered in other
  courses?
\newblock In {\em Proceedings of The 10th International Conference on
  Educational Data Mining (EDM 2017)}, pages 25--28, 2017.

\bibitem{andres16}
J.~M.~L. Andres, R.~S. Baker, G.~Siemens, C.~A. Spann, D.~Ga{\v{s}}evi{\'c},
  and S.~Crossley.
\newblock Studying mooc completion at scale using the mooc replication
  framework.
\newblock 2016.

\bibitem{Brooks2015}
C.~Brooks, C.~Thompson, and S.~Teasley.
\newblock A time series interaction analysis method for building predictive
  models of learners using log data.
\newblock In {\em Proceedings of the Fifth International Conference on Learning
  Analytics And Knowledge}, LAK '15, pages 126--135, New York, NY, USA, 2015.
  ACM.

\bibitem{chen17}
Y.~Chen and M.~Zhang.
\newblock Mooc student dropout: Pattern and prevention.
\newblock In {\em Proceedings of the ACM Turing 50th Celebration Conference -
  China}, ACM TUR-C '17, pages 4:1--4:6, New York, NY, USA, 2017. ACM.

\bibitem{dalgaard2008introductory}
P.~Dalgaard.
\newblock {\em Introductory statistics with R}.
\newblock Springer Science \& Business Media, 2008.

\bibitem{fei15}
M.~Fei and D.-Y. Yeung.
\newblock Temporal models for predicting student dropout in massive open online
  courses.
\newblock In {\em Data Mining Workshop (ICDMW), 2015 IEEE International
  Conference on}, pages 256--263. IEEE, 2015.

\bibitem{gitinabard17w}
N.~Gitinabard, L.~Xue, C.~Lynch, S.~Heckman, and T.~Barnes.
\newblock A social network analysis on blended courses.
\newblock In {\em Proceedings of the 10th International Conference on
  Educational Data Mining, {EDM} 2017(Workshops), Wuhan, China, June 25-28,
  2017}, 2017.

\bibitem{graham06}
C.~R. Graham.
\newblock Blended learning systems.
\newblock {\em The handbook of blended learning}, pages 3--21, 2006.

\bibitem{kloft14}
M.~Kloft, F.~Stiehler, Z.~Zheng, and N.~Pinkwart.
\newblock Predicting mooc dropout over weeks using machine learning methods.
\newblock In {\em Proceedings of the EMNLP 2014 Workshop on Analysis of Large
  Scale Social Interaction in MOOCs}, pages 60--65, 2014.

\bibitem{Kovanovic:Penetrating:2015}
V.~Kovanovic, D.~Gasevic, S.~Dawson, S.~Joksimovic, R.~S. Baker, and M.~Hatala.
\newblock Penetrating the black box of time-on-task estimation.
\newblock In J.~Baron, G.~Lynch, N.~Maziarz, P.~Blikstein, A.~Merceron, and
  G.~Siemens, editors, {\em Proceedings of the Fifth International Conference
  on Learning Analytics And Knowledge, {LAK} '15, Poughkeepsie, NY, USA, March
  16-20, 2015}, pages 184--193. {ACM}, 2015.

\bibitem{kruskal1952use}
W.~H. Kruskal and W.~A. Wallis.
\newblock Use of ranks in one-criterion variance analysis.
\newblock {\em Journal of the American statistical Association},
  47(260):583--621, 1952.

\bibitem{li17}
X.~Li, T.~Wang, and H.~Wang.
\newblock Exploring n-gram features in clickstream data for mooc learning
  achievement prediction.
\newblock In {\em International Conference on Database Systems for Advanced
  Applications}, pages 328--339. Springer, 2017.

\bibitem{munson18}
J.~P. Munson and J.~P. Zitovsky.
\newblock Models for early identification of struggling novice programmers.
\newblock In {\em Proceedings of the 49th ACM Technical Symposium on Computer
  Science Education}, pages 699--704. ACM, 2018.

\bibitem{park2017}
J.~Park, K.~Denaro, F.~Rodriguez, P.~Smyth, and M.~Warschauer.
\newblock Detecting changes in student behavior from clickstream data.
\newblock In {\em Proceedings of the Seventh International Learning Analytics
  \& Knowledge Conference}, LAK '17, pages 21--30, New York, NY, USA, 2017.
  ACM.

\bibitem{pursel2016}
B.~Pursel, L.~Zhang, K.~Jablokow, G.~Choi, and D.~Velegol.
\newblock Understanding mooc students: Motivations and behaviours indicative of
  mooc completion.
\newblock {\em J. Comp. Assist. Learn.}, 32(3):202--217, June 2016.

\bibitem{romero2013web}
C.~Romero, P.~G. Espejo, A.~Zafra, J.~R. Romero, and S.~Ventura.
\newblock Web usage mining for predicting final marks of students that use
  moodle courses.
\newblock {\em Computer Applications in Engineering Education}, 21(1):135--146,
  2013.

\bibitem{ruizpredicting}
S.~Ruiz, M.~Urretavizcaya, and I.~Fern{\'a}ndez-Castro.
\newblock Predicting students' outcome by interaction monitoring.

\bibitem{seaton2014does}
D.~T. Seaton, Y.~Bergner, I.~Chuang, P.~Mitros, and D.~E. Pritchard.
\newblock Who does what in a massive open online course?
\newblock {\em Communications of the ACM}, 57(4):58--65, 2014.

\bibitem{sinha14}
T.~Sinha, N.~Li, P.~Jermann, and P.~Dillenbourg.
\newblock Capturing" attrition intensifying" structural traits from didactic
  interaction sequences of mooc learners.
\newblock {\em arXiv preprint arXiv:1409.5887}, 2014.

\bibitem{Spacco15}
J.~Spacco, P.~Denny, B.~Richards, D.~Babcock, D.~Hovemeyer, J.~Moscola, and
  R.~Duvall.
\newblock Analyzing student work patterns using programming exercise data.
\newblock In {\em Proceedings of the 46th ACM Technical Symposium on Computer
  Science Education}, SIGCSE '15, pages 18--23, New York, NY, USA, 2015. ACM.

\bibitem{willman15}
S.~Willman, R.~Lind{\'e}n, E.~Kaila, T.~Rajala, M.-J. Laakso, and T.~Salakoski.
\newblock On study habits on an introductory course on programming.
\newblock {\em Computer Science Education}, 25(3):276--291, 2015.

\bibitem{zafra12}
A.~Zafra and S.~Ventura.
\newblock Multi-instance genetic programming for predicting student performance
  in web based educational environments.
\newblock {\em Applied Soft Computing}, 12(8):2693--2706, 2012.

\end{thebibliography}

\end{document}